\documentclass[12pt]{article}
\usepackage{times}

\usepackage{epsf}
\def\beq{\begin{equation}}
\def\eeq{\end{equation}}
\def\bea{\begin{eqnarray}}
\def\eea{\end{eqnarray}}

\def\nnb{\nonumber}

\def\rar{\rightarrow}
\def\nnb{\nonumber}

\def\ba{\begin{array}}
\def\ea{\end{array}}
\def\bea{\begin{eqnarray}}
\def\eea{\end{eqnarray}}

\def\BXdtt{$B \rightarrow X_d\, \tau^+ \tau^-$}
\def\BXdll{$B \rightarrow X_d\, \ell^+ \ell^-$}

\title{$B\rightarrow X_d \ell^+\ell^-$  in a CP softly broken  two Higgs doublet model}
\author{\vspace{1cm}\\
         {\bf Hilal Acar}
        \, \, and \, \,
        {\bf G\"{u}rsevil  Turan} \thanks{E-mail address:
        gsevgur@metu.edu.tr}\\Middle East Technical University Physics Dept. Inonu Bul. \\
06531 Ankara-TURKEY}
        \date{}
\begin{document}
\setlength{\baselineskip}{24pt} \maketitle
\setlength{\baselineskip}{7mm}

\abstract{We study the differential branching ratio, forward-backward asymmetry,
CP-violating asymmetry, CP-violating asymmetry
in the forward-backward asymmetry and polarization asymmetries of the final lepton in the \BXdll
decays in the context of  a CP
softly broken two Higgs doublet model. We analyze
the dependencies  of these observables on the model parameters by paying a special
attention to the effects of neutral Higgs boson (NHB) exchanges and possible CP violating
effects. We find that NHB effects are quite significant for the $\tau$ mode. The
above-mentioned observables seems to be  promising
as  a testing ground for new physics beyond the SM, especially for the existence
of the CP-violating phase in the theory.}
\\


 \thispagestyle{empty} \setcounter{page}{1}
\section{Introduction \label{s1}}
Although CP violation is one of the most  fundamental phenomena in particle physics
it is still one of the  the least  tested aspects of the the Standard Model (SM).
Before the start of the B factories, CP violation has only been measured
in the kaon system. Very recently, the  observation of CP violation in the B-meson
system have been reported by the $e^+e^-$ B factories \cite{Bfac} providing the the first test of the SM CP
violation. In the near future, more experimental tests will be possible  at the B factories and possible
deviations from the SM predictions will provide important clues about physics beyond it. This situation
makes the search for CP violation in B decays highly interesting.

Interest in CP violation is not limited to  particle physics; it plays  an important role in
cosmology, too. One of the necessary conditions to generate the
matter-antimatter asymmetry observed in the Universe is -in addition to baryon number violation and deviations
from the thermal equilibrium- that the elementary interactions have to violate CP.
In the SM the only source of CP violation
is the complex  Cabibbo-Kobayashi-Maskawa (CKM) matrix elements which  appears too weak to drive such an
asymmetry \cite{Cohen}, giving a strong motivation to search for new physics. In many cases, extensions
of the SM such as the Two Higgs Doublet Model (2HDM) or the supersymmetric extensions of the SM are able
to supply the new sources of CP violation, providing an opportunity to investigate the new physics by
analyzing the CP violating effects.

Being a FCNC process, $B \rightarrow X_{s,d}\, \ell^+ \ell^-$ decays provide the most reliable testing grounds for the SM at the loop level
and they are also sensitive to new physics. In addition, \BXdll mode is especially important in the CKM
phenomenology. In case of the $b \rar s \ell^+ \ell^-$ decays, the
matrix element receives a combination of various contributions
from the intermediate  $t$, $c$ or $u$ quarks with factors
$V_{tb}V^*_{ts}\sim \lambda^2 $, $V_{cb}V^*_{cs}\sim \lambda^2$
and $V_{ub}V^*_{us}\sim \lambda^4$, respectively, where $\lambda =\sin \theta_C
\cong 0.22$. Since the last factor is extremely
small compared to the other two we can neglect it and this  reduces  the unitarity relation
for the CKM factors  to the form
$V_{tb}V^*_{ts}+V_{cb}V^*_{cs}\approx 0$. Hence, the matrix
element for the $b \rightarrow s\ell^+\ell^-$ decays involve only
one independent CKM factor so that  CP violation would not
show up. On the other hand, as pointed out before \cite{Kruger,Choudhury1}, for $b
\rightarrow d \ell^+\ell^-$ decay, all the CKM factors
$V_{tb}V^*_{td}$, $V_{cb}V^*_{cd}$ and $V_{ub}V^*_{ud}$ are at the
same order $\lambda^3$ in the SM  and the matrix element for these
processes would have sizable interference terms, so as to  induce a
CP violating asymmetry between the decay rates of the reactions $b
\rightarrow d \ell^+\ell^-$ and $\bar{b}\rightarrow
\bar{d}\ell^+\ell^-$. Therefore, $b \rightarrow d \ell^+\ell^-$ decays
seem to be suitable for establishing CP violation in B mesons.

We note that the inclusive $B \rar X_{s} \ell^+ \ell^-$
decays have been widely studied in the framework of the SM and its
various extensions \cite{Hou}-\cite{Ergur}. As for $B \rar X_{d} \ell^+ \ell^-$ modes,
they were first considered within the SM  in \cite{Kruger}
and \cite{Choudhury1}.
The general two Higgs doublet model contributions and minimal supersymmetric extension of
the SM (MSSM) to the CP asymmetries were discussed in refs. \cite{Aliev1}
and \cite{Choudhury2}, respectively.  Recently, CP violation
in the polarized $b \rar d \ell^+ \ell^-$ decay has been also
investigated in the SM \cite{Babu} and also in a general model
independent way \cite{Aliev2}.

The aim of this work is to investigate \BXdll
decay with emphasis on CP violation and NHB effects in a CP softly broken 2HDM,
which is called model IV in the literature \cite{Huang,Huang2}.
In model IV, up-type quarks get masses from Yukawa couplings to the one Higgs doublet,
and down-type quarks and leptons get masses from another Higgs doublet.
In such  a  2HDM, all the parameters in the Higgs potential
are real so that it is CP-conserving, but one allows  the real and imaginary
parts of $\phi^+_1\phi_2$ to have different self-couplings so that the phase $\xi$,
which comes from the expectation value of Higgs field,  can not be rotated away,
which breaks the CP symmetry (for details, see ref \cite{Huang}). In model IV,
interaction vertices of the Higgs bosons and the down-type quarks and leptons depend on
the CP violating phase $\xi$  and the ratio $\tan \beta =v_2/v_1$, where
$v_1$ and $v_2$ are the vacuum expectation values of the first and the second Higgs doublet
respectively, and they are free parameters in the model.  The constraints on $\tan \beta $
are usually obtained from $B-\bar{B}$, $K-\bar{K}$ mixing, $b\rightarrow s \, \gamma$
decay width, semileptonic decay $b\rightarrow c \, \tau \bar{\nu}$ and is given by
\cite{ALEP}
\bea
0.7 \leq \tan \beta \leq 0.52 (\frac{m_{H^{\pm}}}{1 ~ GeV}) \, ,
\eea
and the lower bound $m_{H^{\pm}} \geq 200$ GeV has also been given in \cite{ALEP}.
As for the constraints on $\xi$, it is given in ref.\cite{Huang} that
$\sqrt{|\sin 2 \xi |} \tan \beta < 50$, which can be obtained
from the electric dipole moments of the neutron and electron.

For inclusive B-decays into lepton pairs, in addition to the CP asymmetry and the forward-backward asymmetry,
there is another parameter, namely polarization asymmetry of the final lepton,  which is likely to play
an important role  for comparison of theory with experimental data. It has been already pointed out \cite{ks}
that together with the longitudinal polarization, $P_L$, the other two orthogonal components of polarization,
transverse, $P_T$,  and normal polarizations, $P_N$,
are crucial for the $\tau^+\tau^-$ mode since these three components contain the
independent, but complementary information because they involve different combinations
of Wilson coefficients in addition to the fact that they are proportional to
$m_{\ell}/m_b$.

The paper is organized as follows: Following this  brief introduction,
in section \ref{sect1}, we  first present the effective
Hamiltonian. Then, we introduce  the basic formulas
of the double and differential  decay rates,  CP violation asymmetry, $A_{CP}$,
forward-backward asymmetry, $A_{FB}$, and CP violating asymmetry in
forward-backward asymmetry $A_{CP}(A_{FB})$ for \BXdll decay.
Section \ref{sect2} is devoted to the numerical analysis and discussion.
\section{The Effective hamiltonian for  $B \rar X_{d} \ell^+ \ell^-$ }\label{sect1}
It is well known that
inclusive decay rates of the heavy hadrons can be calculated in the heavy quark effective theory
(HQET) \cite{HQET} and the important result from this procedure is that the leading terms in
$1/m_q$ expansion turn out to be the decay of a free quark, which can be calculated in the
perturbative QCD.
On the other hand, the effective Hamiltonian method provide a  powerful framework for both the
inclusive and the exclusive modes
into which the perturbative QCD corrections to the physical decay amplitude  are incorporated
in a systematic way. In this approach, heavy degrees of freedom,
namely $t$ quark and $W^{\pm}, H^{\pm}, h^0, H^0$ bosons in the present case, are integrated out. The procedure
is to take into account the QCD corrections through matching the full theory with the
effective low energy one at the high scale $\mu =m_W$ and evaluating the Wilson coefficients from $m_W$
down to the lower scale $\mu \sim {\cal O}(m_b)$.
The effective Hamiltonian obtained in this way for the
process $b \rar d \,  \ell^+ \ell^-$, is given by
\cite{Dai,Huang3}:
\begin{eqnarray}\label{Hamiltonian} {\cal H}_{eff} & = &  \frac{4
G_F \alpha}{\sqrt{2}} \, V_{tb} V^*_{td}\Bigg\{ \sum_{i=1}^{10}\, C_i (\mu ) \, O_i(\mu)+
\sum_{i=1}^{10}\, C_{Q_i} (\mu ) \, Q_i(\mu) \nnb\\
& - & \lambda_u \{C_1(\mu)[O_1^u(\mu)-O_1(\mu)]+C_2(\mu)[O_2^u(\mu)-O_2(\mu)]\}\Bigg\}
\end{eqnarray}
where
\bea\label{CKM}
\lambda_u=\frac{V_{ub}V_{ud}^\ast}{V_{tb}V_{td}^\ast},\label{lamu}
\eea
and we have used  the unitarity of the CKM matrix i.e.,
$V_{tb}V_{td}^\ast+V_{ub}V_{ud}^\ast=-V_{cb}V_{cd}^\ast$. The explicit forms of the operators
$O_i$  can be found in  \cite{Grinstein1}.
$O_1^u$ and $O_2^u$ are the new operators for $b\rightarrow d$ transitions
which are absent in the $b\rightarrow s$ decays and given by
\bea
O_1^u & = & (\bar{d}_{\alpha}\gamma_{mu} P_L u_{\beta}) (\bar{u}_{\beta}\gamma^{mu} P_L d_{\alpha})\nnb \\
O_2^u & = & (\bar{d}_{\alpha}\gamma_{mu} P_L u_{\alpha}) (\bar{u}_{\beta}\gamma^{mu} P_L d_{\beta})\nnb .
\eea
The additional operators $Q_i \, (1=1,..,10)$ come from
the NHB exchange diagrams and are defined in ref. \cite{Dai}.

In Eq.(\ref{Hamiltonian}),
$C_i(\mu)$ are the Wilson coefficients calculated at a
renormalization point $\mu$ and their evolution from the higher scale $\mu=m_W$
down to the low-energy scale $\mu=m_b$ is described by the renormalization group
equation. Although  this calculation is performed for operators $O_i$ in the next-to-leading order (NLO)
the mixing of $O_i$ and $Q_i$ in NLO has not been given yet. Therefore we use only the LO results.
The form of the  Wilson coefficients $C_i(m_b)$ and $C_{Q_i}(m_b)$  in the LO are given in refs.
\cite{Grinstein1} and \cite{Dai,Huang}, respectively.

We here present the expression for $C_9(\mu)$ which contains, as well as  a perturbative
part,   a part coming from long distance (LD) effects due to conversion of the
real $\bar{c}c$ into lepton pair $\ell^+ \ell^-$:
\begin{eqnarray}
C_9^{eff}(\mu)=C_9^{pert}(\mu)+ Y_{reson}(s)\,\, ,
\label{C9efftot}
\end{eqnarray}
where
\begin{eqnarray}\label{Cpert}
C_9^{pert}(\mu)&=& C_{9}+h(u,s) [ 3 C_1(\mu) + C_2(\mu) + 3
C_3(\mu) + C_4(\mu) + 3 C_5(\mu) + C_6(\mu) \nonumber
\\&+&\lambda_u(3C_1 + C_2) ] -  \frac{1}{2} h(1, s) \left( 4
C_3(\mu) + 4 C_4(\mu)
+ 3 C_5(\mu) + C_6(\mu) \right)\nnb \\
&- &  \frac{1}{2} h(0,  s) \left[ C_3(\mu) + 3 C_4(\mu) +\lambda_u
(6 C_1(\mu)+2C_2(\mu)) \right] \\&+& \frac{2}{9} \left( 3 C_3(\mu)
+ C_4(\mu) + 3 C_5(\mu) + C_6(\mu) \right) \nonumber \,\, ,
\end{eqnarray}
and
\begin{eqnarray}
Y_{reson}(s)&=&-\frac{3}{\alpha^2}\kappa \sum_{V_i=\psi_i}
\frac{\pi \Gamma(V_i\rightarrow \ell^+
\ell^-)m_{V_i}}{m_B^2 s-m_{V_i}+i m_{V_i}
\Gamma_{V_i}} \nonumber \\
&\times & [ (3 C_1(\mu) + C_2(\mu) + 3 C_3(\mu) + C_4(\mu) + 3
C_5(\mu) + C_6(\mu))\nnb\\ &+&\lambda_u(3C_1(\mu)+C_2(\mu))]\, .
 \label{Yresx}
\end{eqnarray}
In Eq.(\ref{Cpert}), $s=q^2/m_B^2$ where $q$ is the momentum transfer, $u=\frac{m_c}{m_b}$
 and the functions $h(u, s)$ arise from one loop
contributions of the four-quark operators $O_1-O_6$ and are given by
\begin{eqnarray}
h(u, s) &=& -\frac{8}{9}\ln\frac{m_b}{\mu} - \frac{8}{9}\ln u +
\frac{8}{27} + \frac{4}{9} y \\
& & - \frac{2}{9} (2+y) |1-y|^{1/2} \left\{\begin{array}{ll}
\left( \ln\left| \frac{\sqrt{1-y} + 1}{\sqrt{1-y} - 1}\right| -
i\pi \right), &\mbox{for } y \equiv \frac{4u^2}{ s} < 1 \nonumber \\
2 \arctan \frac{1}{\sqrt{y-1}}, & \mbox{for } y \equiv \frac
{4u^2}{ s} > 1,
\end{array}
\right. \\
h(0,s) &=& \frac{8}{27} -\frac{8}{9} \ln\frac{m_b}{\mu} -
\frac{4}{9} \ln s + \frac{4}{9} i\pi \,\, . \label{hfunc}
\end{eqnarray}
The phenomenological parameter $\kappa$
in Eq. (\ref{Yresx}) is taken as $2.3$ (see e.g. \cite{Ligeti}).

Next we proceed to calculate the  differential branching ratio $dBR/ds$,
forward-backward asymmetry $A_{FB}$, CP violating
asymmetry $A_{CP}$, CP asymmetry in the forward-backward asymmetry $A_{CP}(A_{FB})$
and finally the lepton polarization asymmetries    of the \BXdll decays. In order to find these
physically measurable quantities we  first need to calculate the matrix element of the
$B \rar X_{d} \ell^+ \ell^-$ decay.
Neglecting the mass of the $d$ quark,  the effective short distance Hamiltonian
 in Eq.(\ref{Hamiltonian})  leads to the following  QCD corrected matrix element:
\begin{eqnarray}\label{genmatrix}
{\cal M} &=&\frac{G_{F}\alpha}{2\sqrt{2}\pi }V_{tb}V_{td}^{\ast }
\Bigg\{C_{9}^{eff}~\bar{d}\gamma _{\mu }(1-\gamma _{5})b\,\bar{\ell}
\gamma ^{\mu }\ell +C_{10}~\bar{d}\gamma _{\mu }(1-\gamma _{5})b\,\bar{
\ell}\gamma ^{\mu }\gamma _{5}\ell  \nonumber \\
&-&2C_{7}^{eff}~\frac{m_{b}}{q^{2}}\bar{d}i\sigma _{\mu \nu
}q^{\nu }(1+\gamma _{5})b\,\bar{\ell}\gamma ^{\mu }\ell +C_{Q_1}~\bar{d}(1+\gamma _{5})b\,\bar{\ell} \ell
+C_{Q_2}~\bar{d}(1+\gamma _{5})b\,\bar{\ell}\gamma_5 \ell \Bigg\}.
\end{eqnarray}
When the initial and final state polarizations are not measured, we must average over
the initial spins and sum over the final ones, that leads to the following double differential
decay rate
\bea
\frac{d^2 \Gamma}{d s \, dz} & = & \Gamma(B \to X_c \ell \nu) \frac{3\alpha^2 }{4 \pi^2 f(u) k(u)}  (1 - s)^2
    \frac{|V_{tb} V^{\ast}_{td}|^2}{|V_{cb}|^2} \, v \,\Big \{ 2\, v \, z \,
\mbox{\rm Re}(C^{eff}_{7} C^{*}_{10}) \nnb \\
& + & 2 \, \Big(1 + \frac{2 t}{s}\Big ) \,  \mbox{\rm Re}(C^{eff}_{7} C^{eff \,*}_{9})+
 v \,s \, z \, \mbox{\rm Re}(C_{10} C^{eff \,*}_{9}) \nnb \\
& + &  v \sqrt{t}z \, \mbox{\rm Re}((2 C^{eff }_{7}+C^{eff}_{9}) C^{\ast}_{Q_1})+
\sqrt{t}\mbox{\rm Re}(C_{10} C^{\ast}_{Q_2})\nnb \\
& + & \frac{1}{4} \Big [(1+s)-(1-s)\, v^2 z^2+4 t \Big ]|C^{eff }_{9}|^2 +
 \Big [\Big (1+\frac{1}{s}\Big )-\Big (1-\frac{1}{s}\Big )\, v^2 z^2+\frac{4 t}{s}
 \Big ]|C^{eff }_{7}|^2 \nnb \\
& + & \frac{1}{4} \Big [(1+s)-(1-s) \,v^2 z^2-4 t \Big ]|C_{10}|^2+\frac{1}{4} s |C_{Q_2}|^2
+\frac{1}{4} (s-4t) |C_{Q_1}|^2\Big \} \label{DD}
\eea
where $ v=\sqrt{1 - 4 t/s}$, $t=m^2_{\ell}/m^2_b$  and $z=\cos \theta$, where
$\theta$ is the angle between the momentum of the B-meson and that of $\ell^-$
in the center of mass frame of the dileptons $\ell^-\ell^+$. In Eq. (\ref{DD}),
\beq
\Gamma(B \to X_c \ell \nu) = \frac{G_F^2 m_b^5}{192 \pi^3} |V_{cb}|^2 f(u) k(u) \, ,
\eeq
where
\bea
f(u) &=& 1 - 8 u + 8 u^4 - u^8 - 24 u^4 ln (u)   \\
k(u) &=& 1 - \frac{2 \alpha_s(m_b)}{3 \pi}
    \Bigg[ \left( \pi^2 - \frac{31}{4} \right)(1 - \hat{m}_c^2) + {3 \over 2} \Bigg] \, ,
\eea
are the phase space factor and the QCD corrections to the semi-leptonic decay rate, respectively,
which is used to normalize the decay rate of $B \rar X_{d} \ell^+ \ell^-$
to remove the uncertainties in the value of $m_b$.

Having established the double differential decay rates, let us now consider the
forward-backward asymmetry  $A_{FB}$ of the lepton pair, which is defined as
\begin{eqnarray}
A_{FB}(s)& = & \frac{ \int^{1}_{0}dz \frac{d^2 \Gamma }{ds dz} -
\int^{0}_{-1}dz \frac{d^2 \Gamma }{ds dz}}{\int^{1}_{0}dz
\frac{d^2 \Gamma }{ds dz}+ \int^{0}_{-1}dz \frac{d^2 \Gamma }{ds
dz}}~~. \label{AFB1}
\end{eqnarray}
The $A_{FB}$'s for the \BXdll  decays are calculated to be
\begin{eqnarray}
A_{FB}(s) =\frac{-3~ v }{\Delta (s)} \,\mbox{\rm Re}[C_{10}(2 C^{eff}_{7}+s \, C^{eff \,*}_{9})]+
\sqrt{t}\, \mbox{\rm Re}[C_{Q_1}(2 C^{eff\,*}_{7}+  C^{eff \,*}_{9})],
\label{AFB2}
\end{eqnarray}
where
\bea
\Delta (s) & = & \frac{(s+2 s^2+2 t-8 s t)}{s}|C_{10}|^2+\frac{4}{s^2}(2+s)(s+2 t)|C^{eff}_{7}|^2
+(1+2 s)(1+ \frac{2 t}{s})|C^{eff}_{9}|^2 \nnb \\
&+ & \frac{12}{s}(s+2 t)\mbox{\rm Re}(C^{eff}_{7} C^{eff \,*}_{9})+
6 \sqrt{t} \mbox{\rm Re}(C^{eff }_{9}C^{\ast}_{Q_2})+ \frac{3}{2}(s-4 t) |C_{Q_1}|^2
+\frac{3}{2}s |C_{Q_2}|^2 ,\nnb \\ &&
\eea
which agrees with the result given by ref. \cite{Choudhury1}, in case of switching off the NHB
contributions and setting $m_{\ell}=0$,  but differs slightly from the results of  \cite{Choudhury2}.

We next consider the CP  asymmetry $A_{CP}$ between  the $B \rar X_{d} \ell^+ \ell^-$
and the conjugated one $\bar{B} \rar \bar{X_{d}} \ell^+ \ell^-$, which is defined as
\beq
A_{CP}(s) ~=~ \frac{\frac{d \Gamma}{ds} - \frac{d
\bar{\Gamma}}{d s}}{\frac{d \Gamma}{d s} + \frac{d
\bar{\Gamma}}{d s}}
\label{ACPdef}
\eeq
where
\beq
\frac{d\Gamma}{d s} ~=~ \frac{d\Gamma(B \rar X_{d} \ell^+ \ell^-)}{d s} ~~,~~
\frac{d\bar{\Gamma}}{d s} ~=~ \frac{d\Gamma(\bar{B} \rar \bar{X_{d}} \ell^+ \ell^-)}{d s} \,.
\eeq

After integrating the double differential decay rate in Eq.(\ref{DD}) over the angle
variable, we find for the \BXdll decays
\beq
\frac{d \Gamma}{d s}
  = \Gamma(B \to X_c \ell \nu) \frac{\alpha^2 }{4 \pi^2 f(u) k(u)}  (1 - s)^2
    \frac{|V_{tb} V^{\ast}_{td}|^2}{|V_{cb}|^2} \sqrt{1 - \frac{4 t}{s}} \, \Delta (s)\, .
\label{rate}
\eeq
For the antiparticle channel, we have
\bea
\frac{d\bar{\Gamma}}{d s} ~=~ \frac{d\Gamma}{d s} (\lambda_u \rightarrow \lambda^{\ast}_u;\xi \rightarrow -\xi)
\eea

We have also a CP violating asymmetry in $A_{FB}$, $A_{CP}
(A_{FB})$, in \BXdll decay.  Since in the limit of CP conservation, one expects $A_{FB}=-\bar{A}_{FB}$
\cite{Choudhury1,Buch}, where
$A_{FB}$ and $\bar{A}_{FB}$ are the  forward-backward asymmetries in the particle and
antiparticle channels, respectively, $A_{CP}(A_{FB})$ is defined as
\begin{eqnarray}
A_{CP}(A_{FB})& = & A_{FB} +\bar{A}_{FB} ~~, \label{ACPAFB1}
\end{eqnarray}
with
\bea
\bar{A}_{FB}=  A_{FB}(\lambda_u \rightarrow \lambda_u^\ast; \xi \rightarrow -\xi)\, .
\eea


Finally, we would like to discuss the  lepton
polarization effects for the \BXdll decays. The
polarization asymmetries of the final lepton is defined as
\begin{eqnarray}
P_{n} (s) & = & \frac{(d\Gamma (S_n)/ds)-(d\Gamma
(-S_n)/ds)}{(d\Gamma (S_n)/ds)+(d\Gamma (-S_n)/ds)} \, ,\label{PL}
\end{eqnarray}
for $n=L,~N,~T$. Here, $P_L$, $P_T$ and $P_N$ are the longitudinal,
transversal and normal polarizations, respectively. The unit vectors
$S_n$ are defined as follows:
\bea
S_L&=&(0,\vec{e}_L)=\Bigg(0,\frac{\vec{p}_-}{|\vec{p}_-|}\Bigg) \, ,\nnb\\
S_N&=&(0,\vec{e}_N)=\Bigg(0,\frac{\vec{p}\times\vec{p}_-}{|\vec{p}
\times\vec{p}_-|}
\Bigg) \, ,\nnb\\
S_T&=&(0,\vec{e}_T)=\Bigg(0,\vec{e}_N\times
\vec{e}_L \Bigg)\,,\eea
where $\vec{p}$ and $\vec{p}_-$ are  the three-momenta of $d$ quark and $\ell^-$ lepton, respectively.
The longitudinal unit vector
$S_L$ is boosted to the CM frame of $\ell^{+}\ell^{-}$ by Lorentz
transformation: \bea
S_{L,CM}=\Bigg(\frac{|\vec{p}_-|}{m_\ell},\frac{E_\ell~
\vec{p}_-}{m_\ell|\vec{p}_-|}\Bigg).
\eea
It follows from the definition of unit vectors $S_n$ that $P_T$
lies in the decay plane while $P_N$ is perpendicular to it, and
they are not changed by the boost.

After some algebra, we obtain the following expressions
for the polarization components of the $\ell^-$ lepton in \BXdll decays:
\bea
P_{L}&=&\frac{v }{\Delta}\mbox{\rm Re}\Big[ 2 C_{10} (6\, C^{eff,\ast }_{7}+
(1+2 s) C^{eff,\ast }_{9}) -3 C_{Q_1} (2\, \sqrt{t} \, C_{10}+
s C^{\ast }_{Q_2})\Big]\, , \nnb \\
P_{T}&=&\frac{3\pi\sqrt{t}}{2\sqrt{s}\Delta}\Bigg( -\frac{4}{s}|C^{eff }_{7}|^2
-s |C^{eff }_{9}|^2 +\mbox{\rm Re}\Big[2 C^{eff \ast }_{7}(C_{10}-2C^{eff \ast }_{9}+
\frac{s}{2\sqrt{t}}C^{\ast }_{Q_2}) \nnb \\
& + & C^{eff }_{9}(C_{10}+\frac{s}{2\sqrt{t}}C^{\ast }_{Q_2})+
\frac{s-4 t}{2\sqrt{t}} C_{10}C^{\ast }_{Q_1} \Big]\Bigg)\, ,  \\
P_{N}&=&\frac{3\pi v}{4\sqrt{s}\Delta}\mbox{\rm Im}\Big[C_{10}(s C^{\ast }_{Q_2}+
2\sqrt{t} (C^{eff \ast }_{7}+sC^{eff \ast }_{9} )) +s C_{Q_1}(2C^{eff \ast }_{7}+C^{eff \ast }_{9}) \Big]\,.
\nnb
\eea


%
\section{Numerical results and discussion \label{sect2}}
In this section we present the numerical analysis of the inclusive
decays \BXdll in model IV. We will  give the results for only $\ell =\tau$
channel,  which demonstrates the NHB effects more manifestly. The input parameters
we used in this analysis are as follows:
\begin{eqnarray}
& &  m_b =4.8 \, GeV \, , \,m_c =1.4 \,GeV \, , \, m_t =175 \,GeV \, , \,m_{\tau} =1.78 \, GeV \, ,\nnb \\
& &  BR(B\rightarrow X_c e \bar{\nu}_e)=10.4 \%\, , \, m_{H^{\pm}}=200 \, GeV \, ,
m_{H^{0}}=160\, GeV\, ,\,m_{h^{0}}=115\, GeV \, \nnb \\
& &\alpha^{-1}=1/129 \, , \,G_F=1.17 \times 10^{-5}\, GeV^{-2}  \,  .
\end{eqnarray}

The Wolfenstein parametrization \cite{Wolf} of the CKM factor in Eq. (\ref{lamu}) is given by
\bea
\lambda_u=\frac{\rho(1-\rho)-\eta^2-i\eta}{(1-\rho)^2+\eta^2}+O(\lambda^2)\, ,
\eea
and also
\bea
 \frac{|V_{tb} V^{\ast}_{td}|^2}{|V_{cb}|^2} & = & \lambda^2 [(1-\rho)^2+\eta^2]+
 {\cal O}(\lambda^4)\, .
\eea
The updated fitted values for the  parameters $\rho$ and $\eta$ are given as \cite{AliLunghi}
\bea
\bar{\rho} & = & 0.22 \pm 0.07 \, \, (0.25\pm 0.07)\, ,\nnb \\
\bar{\eta} & = & 0.34 \pm 0.04 \, \, (0.34\pm 0.04)\, , \label{param}
\eea
with (without) including the chiral logarithms uncertainties. In our numerical analysis,
we have used  $(\rho,\,\eta)=(0.25;\,0.34)$ .

The masses of the charged and neutral Higgs bosons, $m_{H^\pm}$, $m_{H^0}$,
and $m_{h^0}$, and  the ratio of the vacuum expectation values
of the two Higgs doublets, $\tan\beta$, remain as free parameters
of the model. The restrictions on $m_{H^\pm}$, and $\tan\beta$ have been
already discussed in section \ref{s1}.  For the masses of the neutral Higgs bosons,
the lower limits are given as $m_{H^0}\geq 115$ GeV and $m_{h^0}\geq 89.9$ GeV in \cite{ALEPH}.

In the following, we give results of our calculations about the dependencies of the
differential branching ratio $dBR/ds$,  forward-backward asymmetry $A_{FB}(s)$, CP violating
asymmetry $A_{CP}(s)$, CP asymmetry in the forward-backward asymmetry $A_{CP}(A_{FB})(s)$
and finally the components of the lepton polarization asymmetries, $P_L(s)$, $P_T(s)$ and
$P_N(s)$,  of the \BXdtt decays on the invariant dilepton mass $s$. In order to
investigate the dependencies of the above physical quantities on the model parameters, namely
CP violating phase $\xi$ and $\tan \beta$, we eliminate the other parameter $s$ by performing
the $s$ integrations over the allowed kinematical region  so as to obtain
their  averaged values, $<A_{FB}>$, $<A_{CP}>$, $<A_{CP}(A_{FB})>$, $<P_L>$, $<P_T>$ and
$<P_N>$.

\begin{figure}[htb]
\vskip 0truein \centering \epsfxsize=3.8in
\leavevmode\epsffile{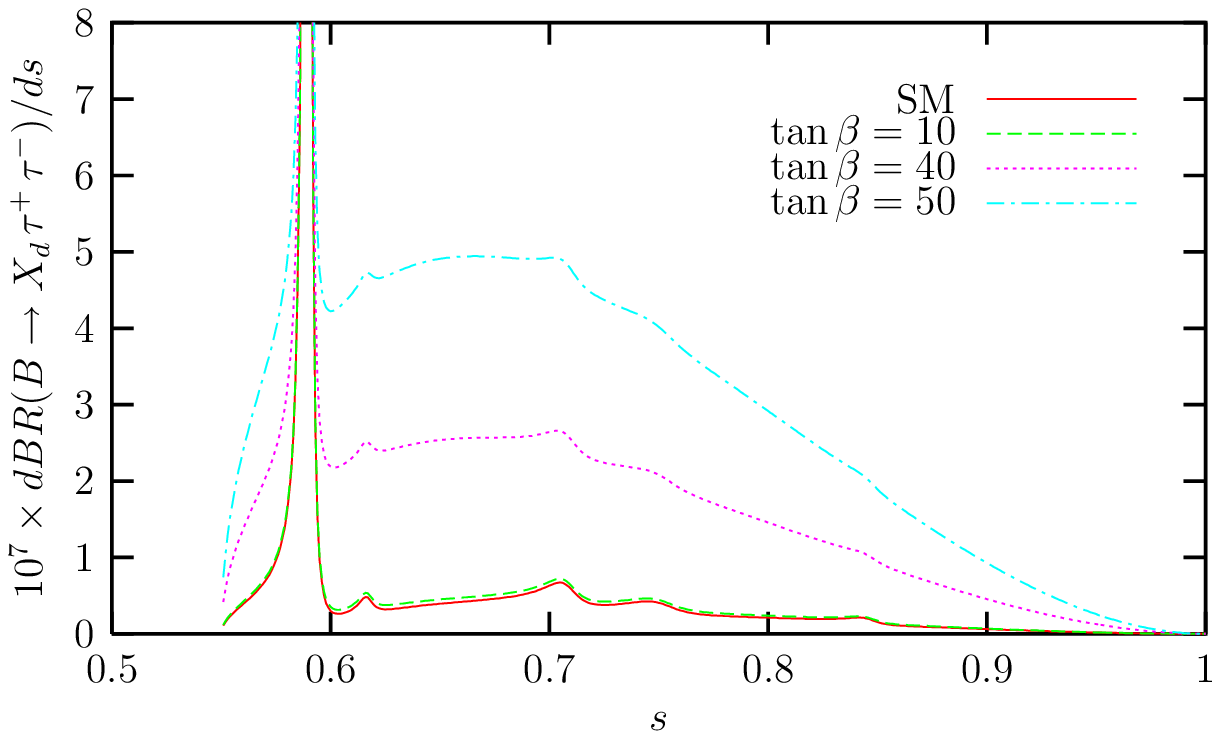} \vskip 0truein \caption[]{Differential branching ratio
as a function of $s$, where $\xi=\pi/4$.} \label{dBR}
\end{figure}

\begin{figure}[htb]
\vskip 0truein \centering \epsfxsize=3.8in \leavevmode
\epsffile{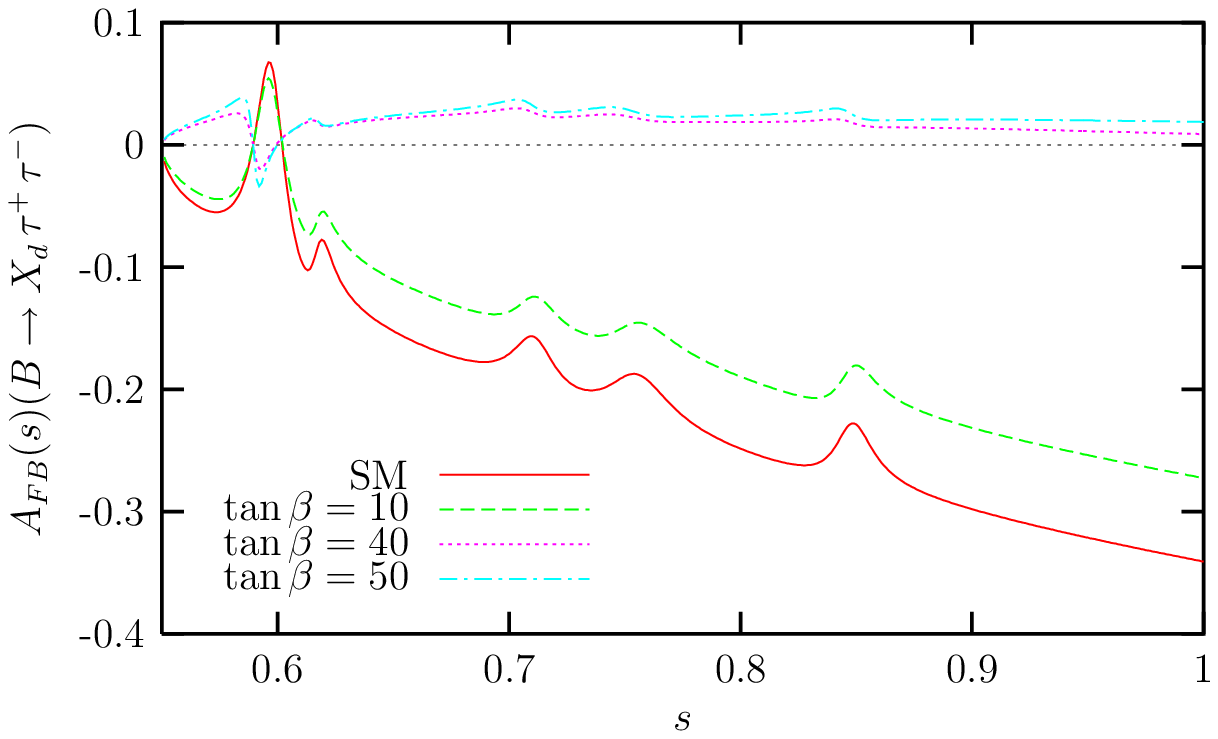} \vskip 0truein \caption{ The forward-backward asymmetry as a
function of $s$, where $\xi=\pi/4$.} \label{dAFB}
\end{figure}

\begin{figure}[htb]
\vskip 0truein \centering \epsfxsize=3.8in
\leavevmode\epsffile{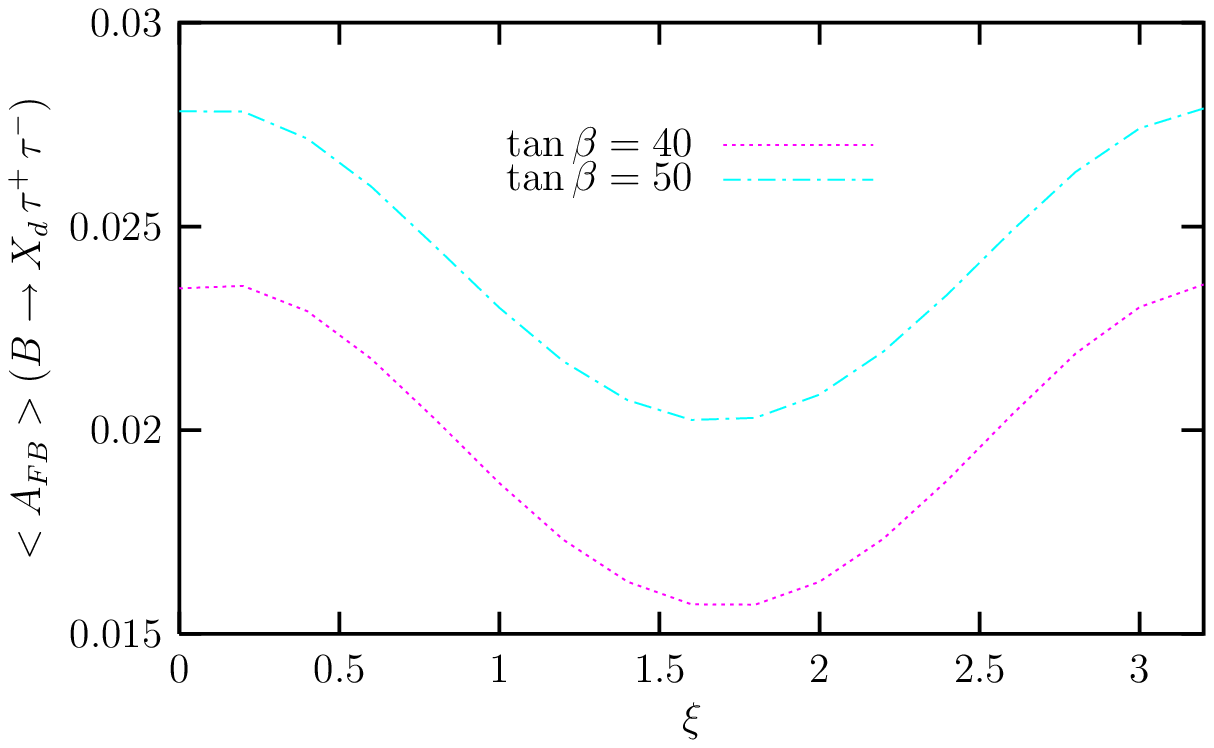} \vskip 0truein \caption{ $<A_{FB}>$  as a function of  $\xi$.}
\label{AFBkcp}
\end{figure}

\begin{figure}[htb]
\vskip 0truein \centering \epsfxsize=3.8in
\leavevmode\epsffile{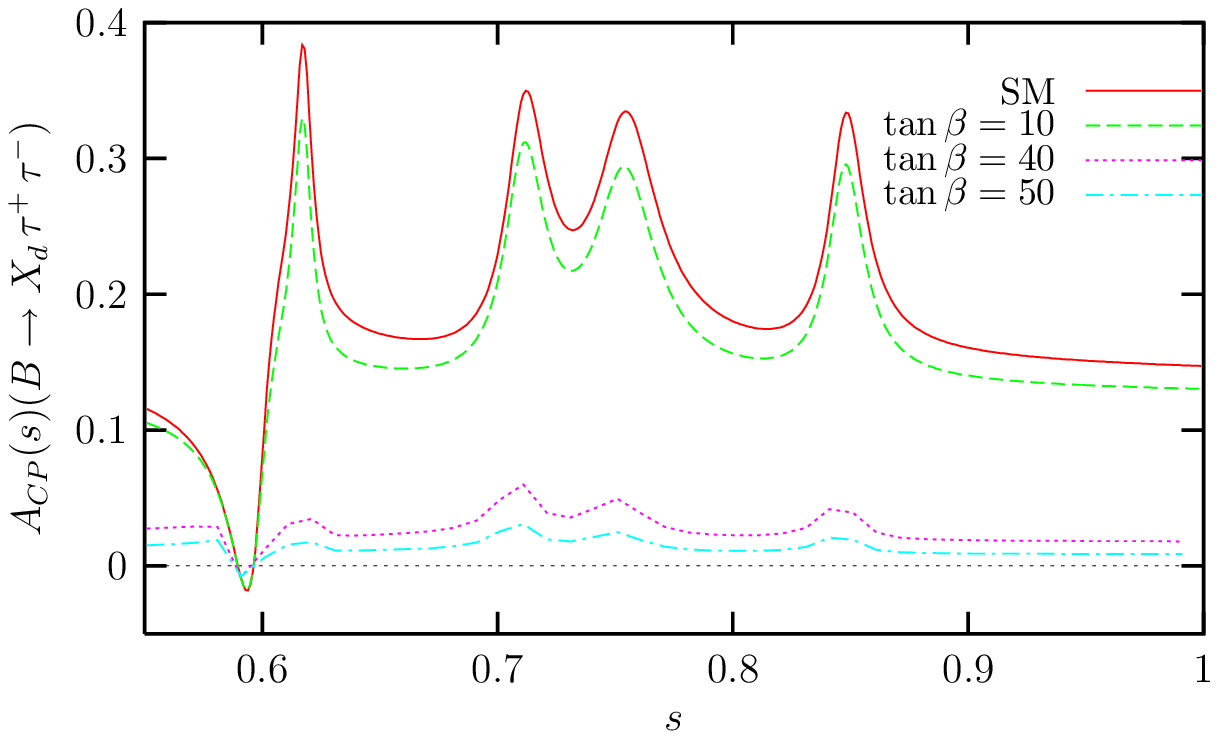} \vskip 0truein \caption{The CP asymmetry as a
function of $s$, where $\xi=\pi /4$.} \label{dACP}
\end{figure}

\begin{figure}[htb]
\vskip 0truein \centering \epsfxsize=3.8in
\leavevmode\epsffile{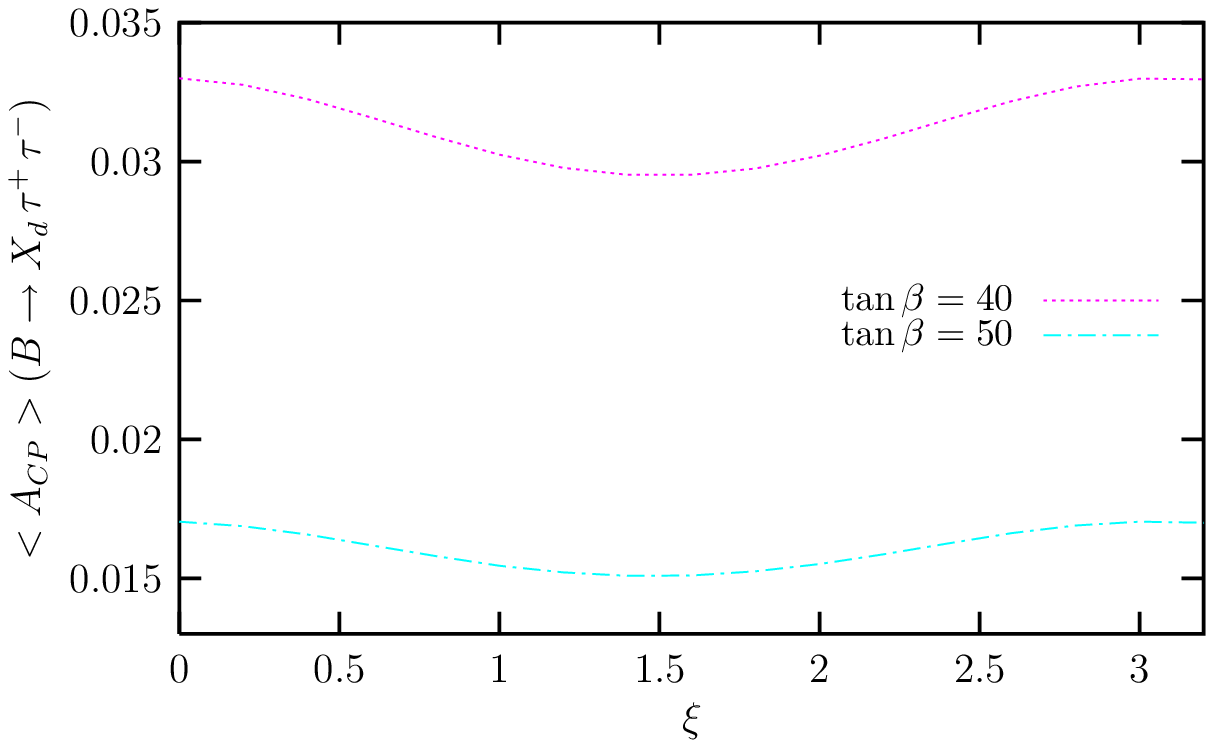} \vskip 0truein \caption{$<A_{CP}>$ as a function of $\xi$.}
\label{ACPkcp}
\end{figure}

\begin{figure}[htb]
\vskip 0truein \centering \epsfxsize=3.8in
\leavevmode\epsffile{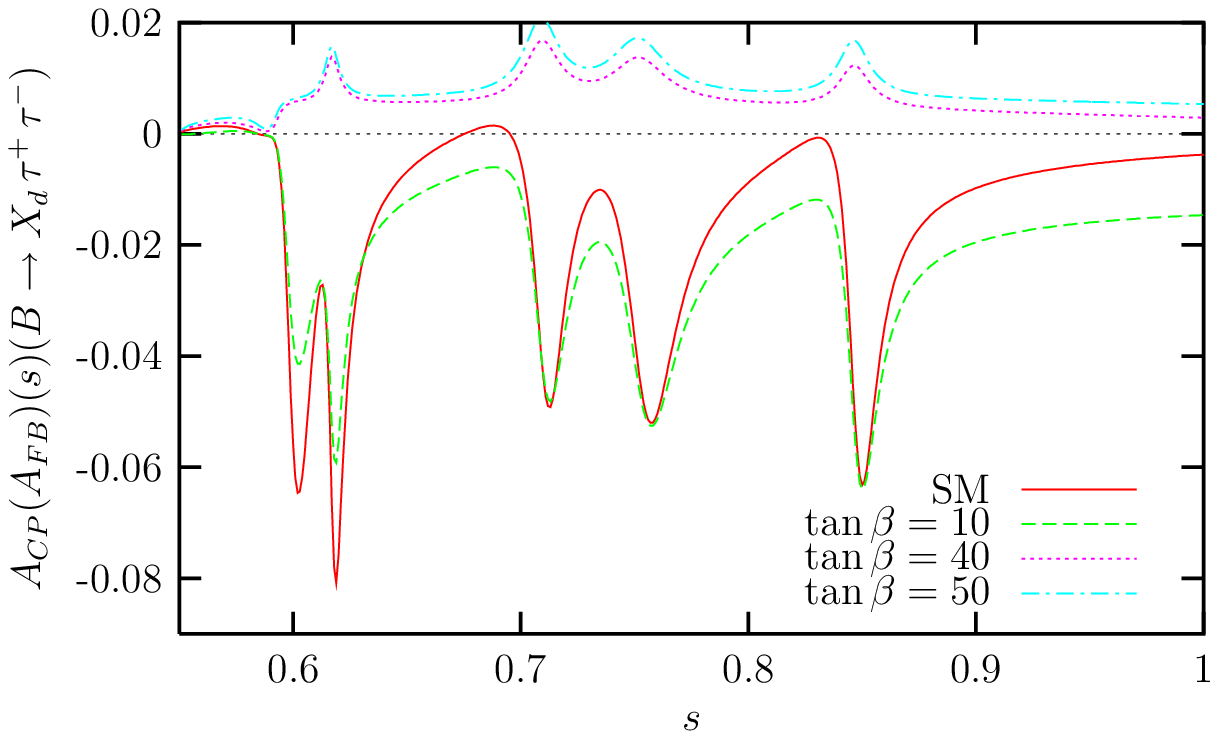} \vskip 0truein \caption{The CP  asymmetry in the
forward-backward asymmetry as  a function of $s$, where $\xi=\pi /4$.}
\label{dACPAFB}
\end{figure}

\begin{figure}[htb]
\vskip 0truein \centering \epsfxsize=3.8in
\leavevmode\epsffile{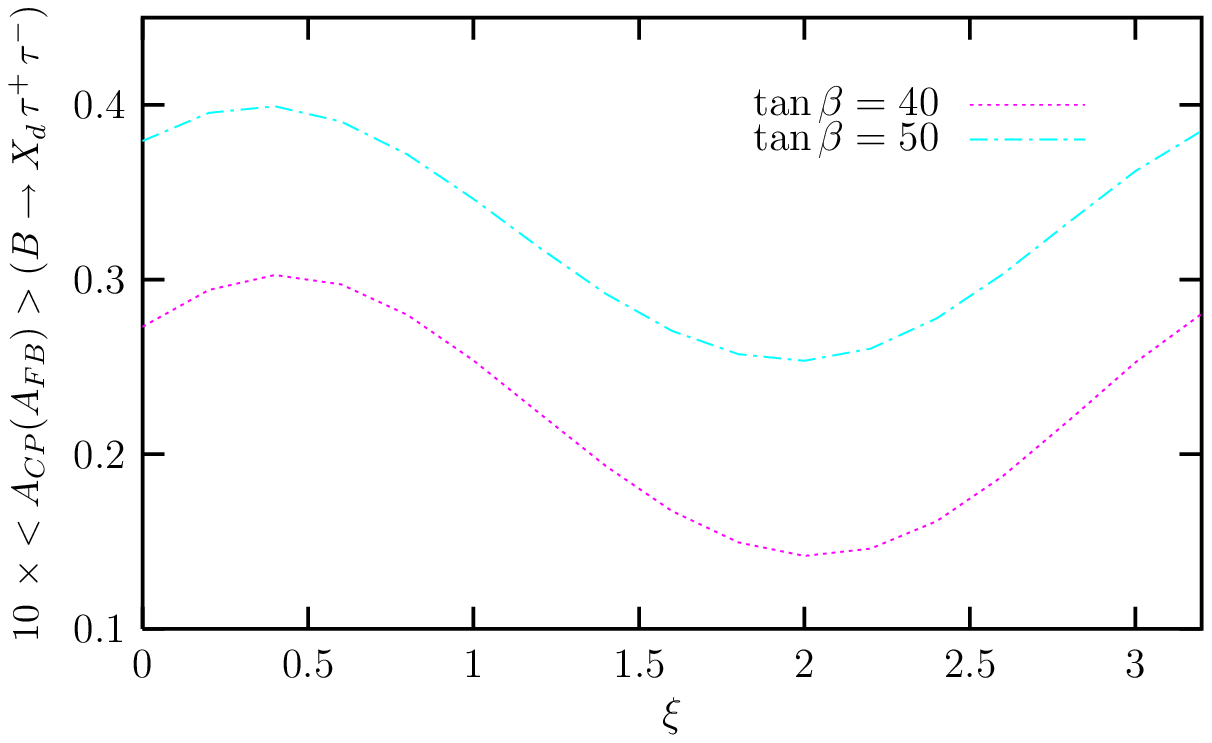} \vskip 0truein \caption{ The CP  asymmetry in the
forward-backward asymmetry as  a function of $\xi$.}
\label{ACPAFBkcp}
\end{figure}

\begin{figure}[htb]
\vskip 0truein \centering \epsfxsize=3.8in
\leavevmode\epsffile{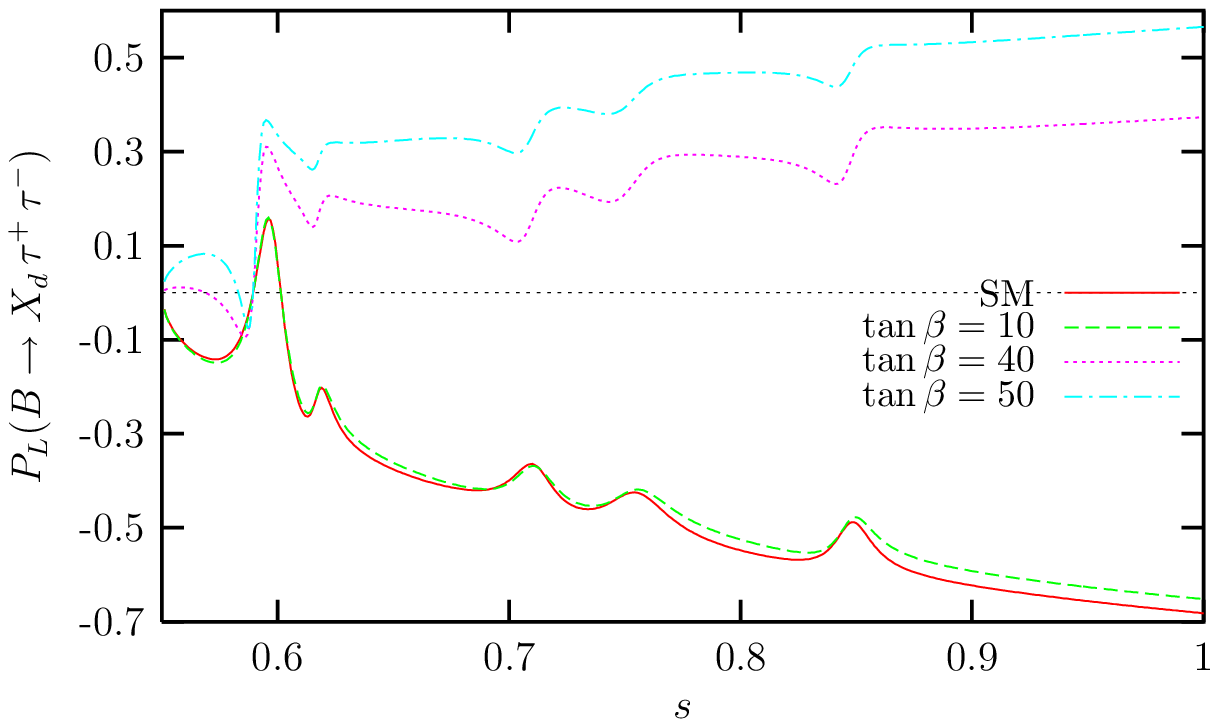} \vskip 0truein \caption{ $P_L (s)$ as a function of  $s$
, where $\xi=\pi /4$. } \label{dPL}
\end{figure}

\begin{figure}[htb]
\vskip 0truein \centering \epsfxsize=3.8in
\leavevmode\epsffile{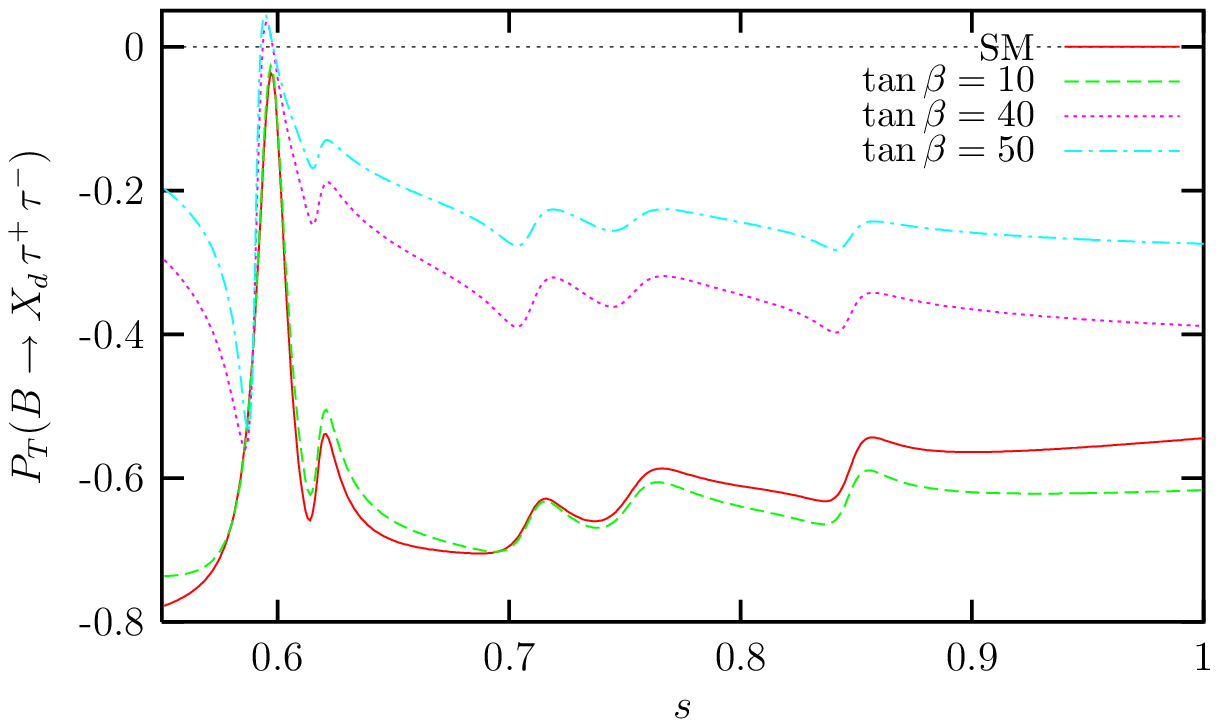} \vskip 0truein \caption{$P_T (s)$ as a function of  $s$
, where $\xi=\pi /4$.}
\label{dPT}
\end{figure}

\begin{figure}[htb]
\vskip 0truein \centering \epsfxsize=3.8in
\leavevmode\epsffile{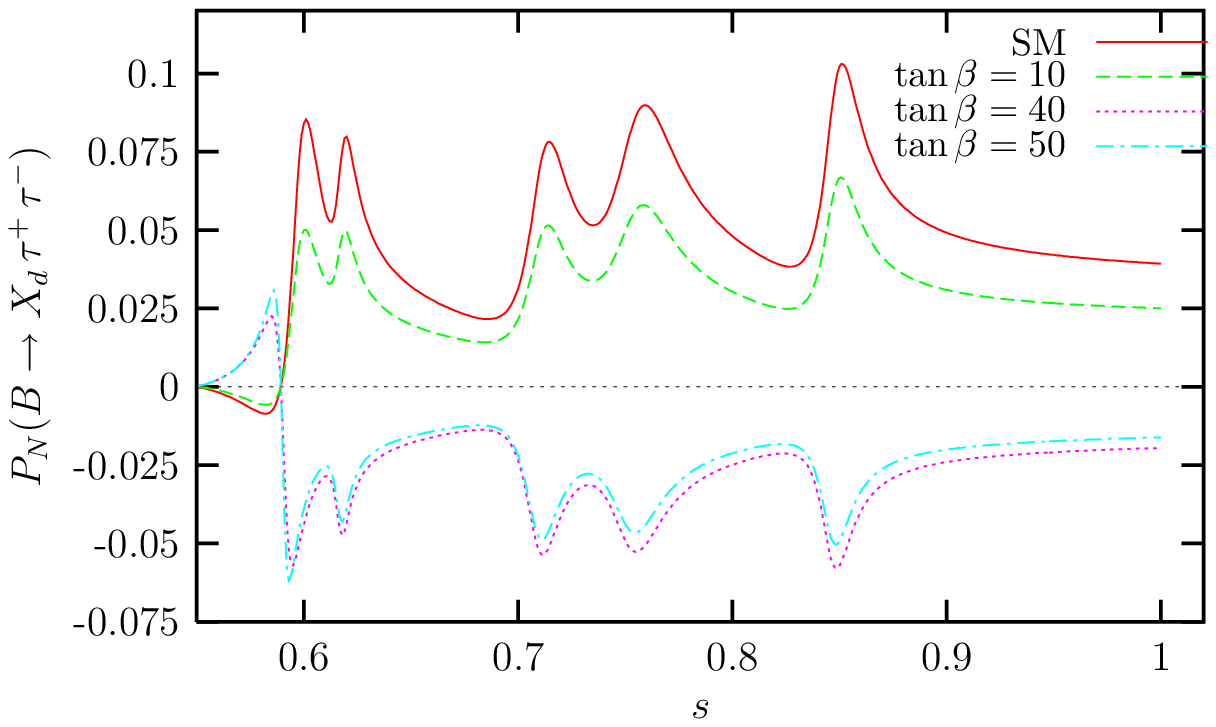} \vskip 0truein \caption{$P_N (s)$ as a function of  $s$
, where $\xi=\pi /4$.}
\label{dPN}
\end{figure}

\begin{figure}[htb]
\vskip 0truein \centering \epsfxsize=3.8in
\leavevmode\epsffile{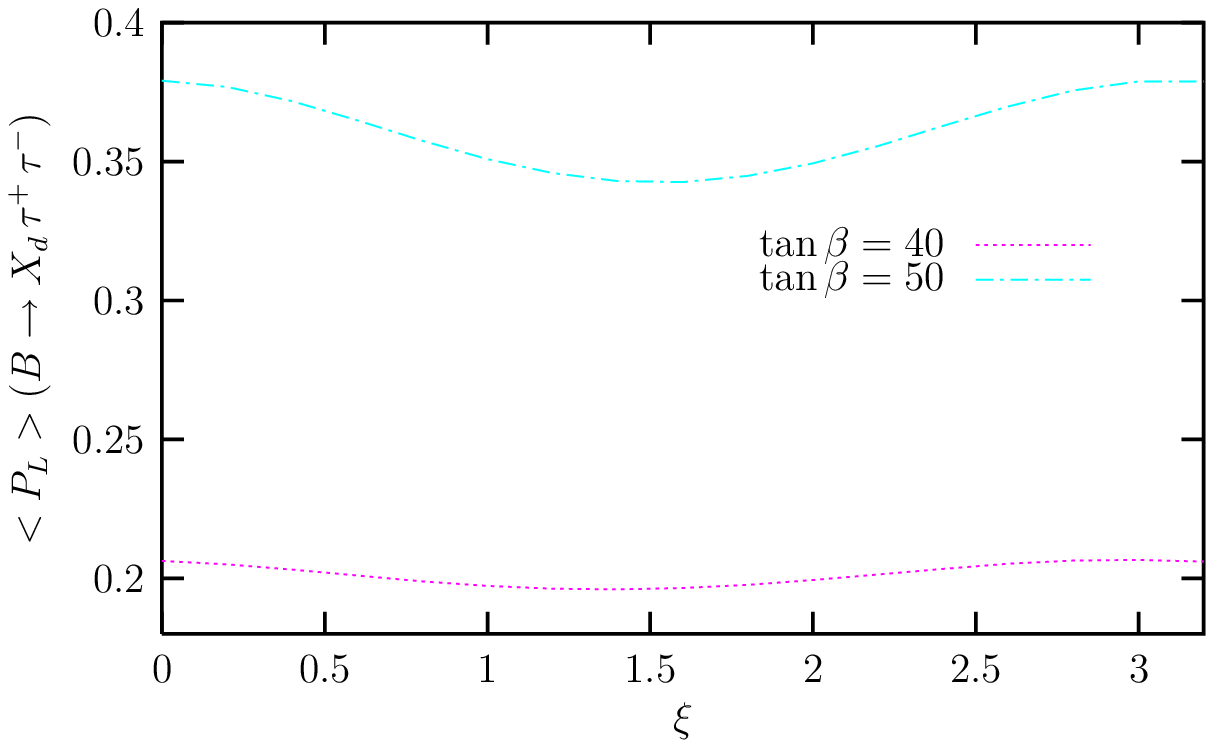} \vskip 0truein \caption{$<P_L >$ as a function of $\xi$. }
\label{PLkcp}
\end{figure}

\begin{figure}[htb]
\vskip 0truein \centering \epsfxsize=3.8in
\leavevmode\epsffile{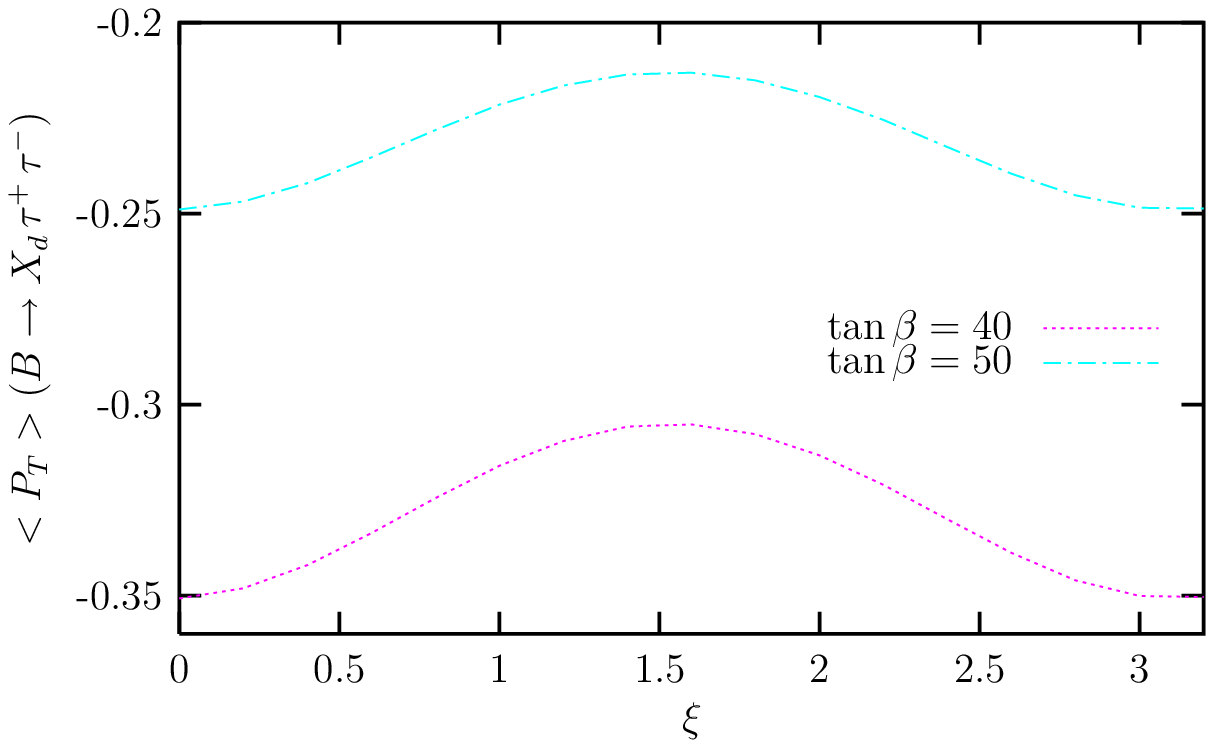} \vskip 0truein \caption{$<P_T >$ as a function of $\xi$. }
\label{PTkcp}
\end{figure}

\begin{figure}[htb]
\vskip 0truein \centering \epsfxsize=3.8in
\leavevmode\epsffile{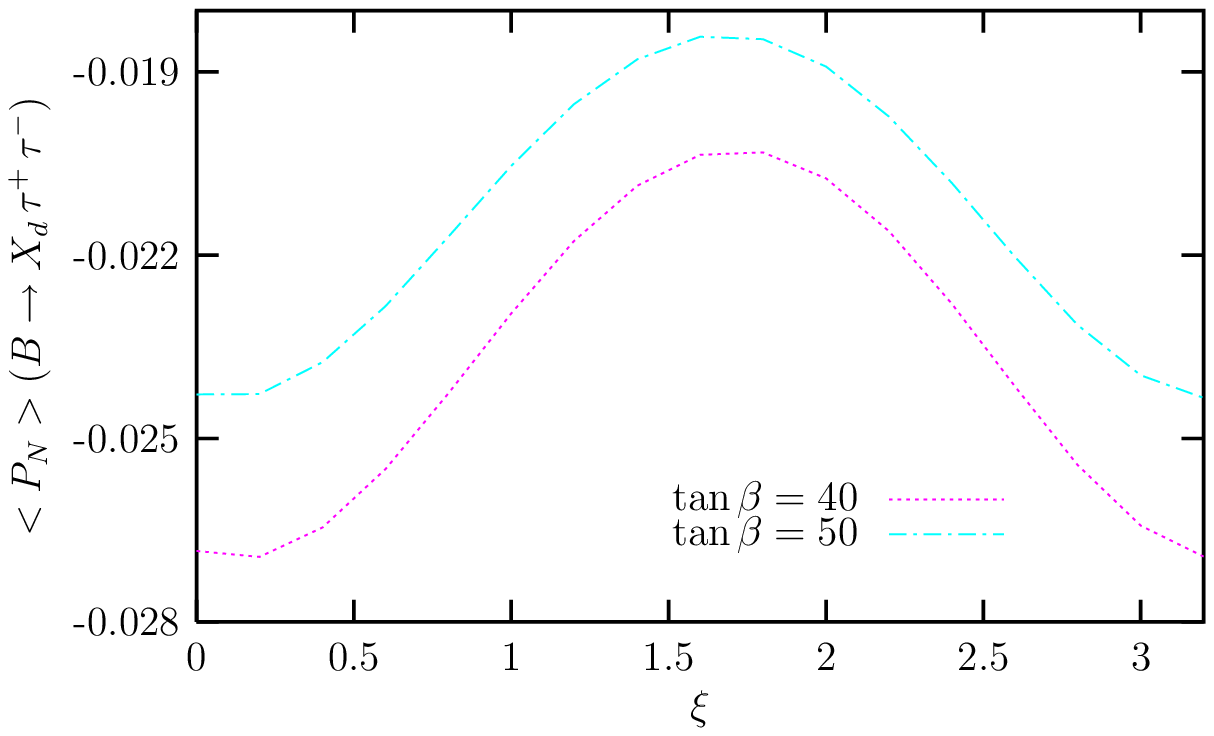} \vskip 0truein \caption{$<P_N >$ as a function of $\xi$. .}
\label{PNkcp}
\end{figure}

Numerical results are shown in Figs. (\ref{dBR})-(\ref{PNkcp}) and
we have the following line conventions: dashed lines, dot lines and dashed-dot lines
represent the model IV contributions with $\tan \beta =10, 40, 50$, respectively and
the  solid lines are for the SM predictions. The cases of switching off NHB contributions
i.e., setting $C_{Q_i}=0$, almost coincide with the cases of 2HDM contributions
with $\tan \beta =10$, therefore we did not plot them separately.

In Fig.(\ref{dBR}), we give  the dependence of the $dBR/ds$ on $s$.
From this figure  NHB effects are very obviously seen,  especially in the moderate-s region.

In Fig. (\ref{dAFB}) and Fig. (\ref{AFBkcp}),  $A_{FB}(s)$ and $<A_{FB}>$  as a
function of $s$ and CP violating phase $\xi$ are presented, respectively.   We see that $A_{FB}$
is more sensitive to $\tan \beta$ than the $dBR/ds$ and it changes sign with the different choices of this
parameter. It is seen from Fig.(\ref{AFBkcp}) that $<A_{FB}>$  is quite sensitive to $\xi$ and
 between $(0.15,0.28)\times 10^{-1}$. We also observe that $<A_{FB}>$ differs essentially from the one
 predicted by the CP-conservative 2HDM (model II, for examples, see \cite{Haber}), which is $0.028$ and $0.023$ for $\tan \beta =40,50$, respectively. In region
 $1<\xi<2$ change in $<A_{FB}>$ with respect to model II reaches $25 \%$.

Figs. (\ref{dACP}) and  (\ref{ACPkcp}) show the dependence of $A_{CP}(s)$
on $s$ and  $<A_{CP}>$  on $\xi$, respectively. We see that $A_{CP}(s)$
is also sensitive to $\tan \beta$ and its sign does not change in the allowed
values of $s$ except in the resonance mass region. It follows from Fig. (\ref{ACPkcp})
that $<A_{CP}>$ is not as  sensitive as $<A_{FB}>$ to $\xi$, and it varies in the range
 $(0.15,0.33)\times 10^{-1}$.

$A_{CP}(A_{FB})(s)$ and $<A_{CP}(A_{FB})>$ of \BXdtt as a
function of $s$ and CP violating phase $\xi$ are presented in Fig. (\ref{dACPAFB}) and
Fig. (\ref{ACPAFBkcp}), respectively. We see that $A_{CP}(A_{FB})(s)$
 changes sign with the different choices of $\tan \beta$.  $<A_{CP}(A_{FB})>$  is between
 $(0.010,0.040)$ and  differs essentially from the one
 predicted by model II, which is $0.038$ and $0.027$ for $\tan \beta =40,50$, respectively. In region
 $1.5<\xi<2.5$ change in $<A_{FB}>$ with respect to model II reaches $35 \%$.

In Figs. (\ref{dPL})-(\ref{dPN}), we present the $s$  dependence of the
longitudinal $P_L$, transverse $P_T$ and normal $P_N$ polarizations of the final lepton
for \BXdtt decay. It is seen that NHB contributions changes the polarization significantly, especially when
$\tan \beta$ is large. We also observe  that except the resonance region, $P_T$ is negative for
all values of $s$, but $P_L$ and
$P_N$ change sign with the different choices of the values of $\tan \beta$.
In Figs. (\ref{PLkcp})-(\ref{PNkcp}),  dependence of the averaged values of the
longitudinal $<P_L>$, transverse $<P_T>$ and normal $<P_N>$ polarizations of the final
lepton for \BXdtt decay  on $\xi$ are shown. It is obvious from these figures that
$<P_N>$ and  $<P_T>$ are more sensitive to $\xi$ than $<P_L>$. In region
 $1.5<\xi<2.0$ change in $<P_N>$  with respect to model II reaches $25 \%$. Thus, measurement of this component
in future experiments may provide information about the model IV parameters.

Therefore, the experimental investigation of $A_{FB}$, $A_{CP}$, $A_{CP}(A_{FB})$ and
the polarization components in \BXdll decays may be quite suitable for  testing the new physics
effects beyond the SM.

\newpage

\end{document}